\documentclass[twocolumn,tighten,times]{aastex631}
\usepackage{amsmath}
\usepackage{anyfontsize}
\usepackage{amssymb}
\usepackage{sidecap}
\usepackage{natbib}
\usepackage{lipsum}
\usepackage{nameref}
\usepackage[T1]{fontenc}
\usepackage{graphicx}
\usepackage{hyperref}

\newcommand{\tess}{\textit{TESS}}
\newcommand{\kepler}{\textit{Kepler}}
\newcommand{\ktwo}{\textit{Kepler\slash K2}}
\newcommand\blfootnote[1]{%
  \begingroup
  \renewcommand\thefootnote{}\footnote{#1}%
  \addtocounter{footnote}{-1}%
  \endgroup
}

\shorttitle{AASTeX v6.3.1 Sample article}
\shortauthors{Wang et al.}
\graphicspath{{./}{figures/}}
\begin{document}

\setlength{\abovedisplayskip}{3pt}
\setlength{\belowdisplayskip}{3pt}

\newcommand{\JHU}{Physics and Astronomy Department, Johns Hopkins University, Baltimore, MD 21218, USA}
\newcommand{\JHUCS}{Department of Computer Science, Johns Hopkins University, Baltimore, MD 21218, USA}
\newcommand{\STScI}{Space Telescope Science Institute, Baltimore, MD 21218, USA}
\newcommand{\NCPAS}{National Centre for the Public Awareness of Science, Australian National University, Canberra, ACT 2611, Australia}
\newcommand{\UCSC}{Department of Astronomy and Astrophysics, University of California, Santa Cruz, CA 95064, USA}
\newcommand{\NSF}{National Science Foundation Graduate Research Fellow}
\newcommand{\CHEIEF}{CHE Israel Excellence Fellowship}
\newcommand{\IESF}{ISEF Fellowship}
\newcommand{\anuobs}{Mt Stromlo Observatory, The Research School of Astronomy and Astrophysics, Australian National University, ACT 2601, Australia}
\newcommand{\canterbury}{School of Physical and Chemical Sciences | Te Kura Matū, University of Canterbury, Private Bag 4800, Christchurch 8140, New Zealand}
\newcommand{\CIERA}{Center for Interdisciplinary Exploration and Research in Astrophysics (CIERA), Northwestern University, Evanston, IL 60208,USA}
\newcommand{\ICECSIC}{Institute of Space Sciences (ICE, CSIC), Campus UAB, Carrer de Can Magrans, s/n, E-08193 Barcelona, Spain}
\newcommand{\IEEC}{Institut d’Estudis Espacials de Catalunya (IEEC), E-08034 Barcelona, Spain}
\newcommand{\astrotd}{The ARC Centre of Excellence for All-Sky Astrophysics in 3 Dimension (ASTRO 3D), Australia}
\newcommand{\anucga}{Centre for Gravitational Astrophysics, College of Science, The Australian National University, ACT 2601, Australia}
\newcommand{\anu}{The Research School of Astronomy and Astrophysics, Australian
National University, ACT 2601, Australia}
\newcommand{\qub}{Astrophysics Research Centre, School of Mathematics and Physics, Queen's University Belfast, Belfast BT7 1NN, UK}
\newcommand{\ESO}{European Southern Observatory, Alonso de C\'ordova 3107, Casilla 19, Santiago, Chile}
\newcommand{\MIAMAS}{Millennium Institute of Astrophysics MAS, Nuncio Monsenor Sotero Sanz 100, Off. 104, Providencia, Santiago, Chile}
\newcommand{\warwick}{Department of Physics, University of Warwick, Gibbet Hill Road, Coventry CV4 7AL, UK}
\newcommand{\VT}{Department of Physics, Virginia Tech, 850 West Campus Drive, Blacksburg VA, 24061, USA}
\newcommand{\cifar}{CIFAR Azrieli Global Scholars program, CIFAR, Toronto, Canada}
\newcommand{\TelAviv}{The School of Physics and Astronomy, Tel Aviv University, Tel Aviv 69978, Isreal}
\newcommand{\birmingham}{Birmingham Institute for Gravitational Wave Astronomy and School of Physics and Astronomy, University of Birmingham, Birmingham B15 2TT, UK}
\newcommand{\OKC}{The Oskar Klein Centre, Department of Astronomy, Stockholm University, AlbaNova, SE-10691 Stockholm, Sweden}
\newcommand{\Uwarsaw}{Astronomical Observatory, University of Warsaw, Al. Ujazdowskie 4, 00-478 Warszawa, Poland}
\newcommand{\ctio}{Cerro Tololo Inter-American Observatory, NSF’s NOIRLab, Casilla 603, La Serena, Chile}
\newcommand{\lco}{Las Cumbres Observatory, 6740 Cortona Dr, Suite 102, Goleta,CA 93117-5575, USA}
\newcommand{\ucsb}{Department of Physics, University of California, Santa Barbara, CA 93106-9530, USA}
\newcommand{\oxf}{Department of Physics, University of Oxford, Oxford, OX1 3RH, UK}
\newcommand{\gemini}{Gemini Observatory, NSF's NOIRLab, 670 N. A'ohoku Place, Hilo, Hawai'i, 96720, USA}
\title{Revealing the progenitor of SN~2021zby through analysis of the \tess\  shock-cooling light curve}

\author[0000-0001-5233-6989]{Qinan~Wang}\blfootnote{Corresponding author: Qinan~Wang\\ \href{mailto:qwang75@jhu.edu}{qwang75@jhu.edu}}
\affiliation{\JHU}
\author[0000-0003-1997-3649]{Patrick~Armstrong}
\affiliation{\anuobs}
\author[0000-0002-0632-8897]{Yossef Zenati}
\altaffiliation{\CHEIEF}
\altaffiliation{\IESF}
\affiliation{\JHU}
\author[0000-0003-1724-2885]{Ryan~Ridden-Harper}
\affiliation{\canterbury}
\author[0000-0002-4410-5387]{Armin Rest}
\affil{\STScI}
\affil{\JHU}
\author[0000-0001-7090-4898]{Iair~Arcavi}
\affil{\TelAviv}
\affil{\cifar}
\author[0000-0002-5740-7747]{Charles~D.~Kilpatrick}
\affil{\CIERA}
\author[0000-0002-2445-5275]{Ryan~J.~Foley}
\affil{\UCSC}
\author[0000-0002-5740-7747]{Brad~E.~Tucker}
\affil{\anuobs}
\affil{\NCPAS}
\affil{\astrotd}
\author[0000-0002-5740-7747]{Chris Lidman}
\affil{\anucga}
\affil{\anu}
\author[0000-0002-0440-9597]{Thomas~L.~Killestein}
\affil{\warwick}
\author[0000-0002-9301-5302]{Melissa Shahbandeh}
\affil{\JHU}

\author[0000-0003-0227-3451]{Joseph P Anderson}
\affil{\ESO}
\affil{\MIAMAS}
\author[0000-0002-5221-7557]{Chris~Ashall}
\affil{\VT}
\author[0000-0003-0035-6659]{Jamison~Burke}
\affil{\lco}
\affil{\ucsb}
\author[0000-0002-1066-6098]{Ting-Wan Chen}
\affil{\OKC}
\author{Kyle A. Dalrymple}
\affil{\JHU}
\author[0000-0002-5680-4660]{Kyle W. Davis}
\affil{\UCSC}
\author[0000-0003-1916-0664]{Michael D. Fulton}
\affil{\qub}
\author[0000-0002-1296-6887]{Lluís Galbany}
\affil{\ICECSIC}
\affil{\IEEC}
\author[0000-0002-1650-1518]{Mariusz Gromadzki}
\affil{\Uwarsaw}
\author{Nada Ihanec}
\affil{\Uwarsaw}
\affil{\ESO}
\author[0000-0001-5754-4007]{Jacob~E.~Jencson}
\affil{\JHU}
\author[0000-0002-6230-0151]{David~O.~Jones}%{David~O.~Jones}
\affiliation{\gemini}
\author[0000-0002-3464-0642]{Joseph~D.~Lyman}
\affil{\warwick}
\author[0000-0003-3939-7167]{Tomás~E.~Müller-Bravo}
\affil{\ICECSIC}
\affil{\IEEC}
\author[0000-0001-9570-0584]{Megan~Newsome}
\affil{\lco}
\affil{\ucsb}
\author[0000-0002-2555-3192]{Matt~Nicholl}
\affil{\birmingham}
\author{David~O'Neill}
\affil{\warwick}
\author[0000-0002-7472-1279]{Craig~Pellegrino}
\affil{\lco}
\affil{\ucsb}
\author[0000-0002-3825-0553]{Sofia~Rest}
\affil{\JHUCS}
\author[0000-0002-8229-1731]{Stephen~J. Smartt}
\affil{\qub}
\affil{\oxf}
\author[0000-0001-9535-3199]{Ken~Smith}
\affiliation{\qub}
\author[0000-0003-4524-6883]{Shubham Srivastav}
\affil{\qub}
\author[0000-0002-1481-4676]{Samaporn~Tinyanont}
\affil{\UCSC}
\author[0000-0002-1229-2499]{David R. Young}
\affil{\qub}
\author[0000-0001-6455-9135]{Alfredo~Zenteno}
\affil{\ctio}

\begin{abstract}

We present early observations and analysis of the double-peaked Type IIb supernova (SN IIb) 2021zby. \tess\ captured the prominent early shock cooling peak of SN~2021zby within the first $\sim$10~days after explosion with a 30-minute cadence.  
We present optical and near-infrared spectral series of SN~2021zby, including three spectra during the shock cooling phase.
Using a multi-band model fit, we find that the inferred properties of its progenitor are consistent with a red supergiant or yellow supergiant, with an envelope mass of $\sim$0.3--3.0~M$_\odot$ and an envelope radius of $\sim$50--350~$R_\odot$. These inferred progenitor properties are similar to those of other SNe~IIb with double-peak feature, such as SNe~1993J, 2011dh, 2016gkg and 2017jgh. This study further validates the importance of the high cadence and early coverage in resolving the shape of the shock cooling light curve, while the multi-band observations, especially UV, is also necessary to fully constrain the progenitor properties. 

%\redpen{add comparison to other SNe}
% to check: 16gkg, 93J, 11dh, others?
% 2016gkg: specra series hard to get. only 

\end{abstract}



\section{Introduction}\label{sec:intro}
%things to discuss:
%\begin{itemize}
%    \item \redpen{Does lightcurve fitting agree with spectra?}
%    \item \redpen{How deep we want to go in spectroscopic analysis?}
%    \item \redpen{any other insights}
%\end{itemize}

Type IIb supernovae (SNe IIb) are characterized by the presence of hydrogen lines at early phases typical of Type II supernovae (SNe II), which fade at later phases as helium features begin to dominate the spectra \citep{filippenko93}. In the weeks following the explosion, the spectra of the supernovae will therefore transition from Type II to Type I. This evolution can be explained by the progenitor star losing most, but not all, of its hydrogen-rich envelope before explosion. 
%This incomplete stripping of hydrogen thus leaves the SN IIb with an early hydrogen signature. 
The exact progenitors of SNe IIb are currently not fully understood with two leading possibilities \citep{EnsmanWoosley1988,Woosley+93, Heger+03, Dessart+2011,SmithHB17,Sravan+20,Long+22}: 1. a low mass star ($<20\ M_\odot$) in a binary system or 2. an isolated high mass star (25--80 $M_\odot$).

The transition from hydrogen-rich progenitors to stripped-envelope progenitors is not fully understood, neither is the exact mass of hydrogen \citep{Gilkis&Arcavi22} as two evolutionary pathways are possible and each scenario involves different masses and nuclear burning instabilities \citep{Arnett&Meakin11,Arnett+18}. In the low mass binary case, mass loss occurs through binary interactions when one of the stars enters its red giant phase \citep{Sana+12,Soker17,yoon17,Gilkis+19,Lohev+19}. In the second case of a high mass star, mass loss is believed to be as a result of strong stellar winds.
%\redpen{(this should have a ref)}
In a few cases, the progenitors of SNe IIb have been  directly identified in pre-SN images as supergiants with radii $\gtrsim 200R_\odot$, such as SN~1993J \citep{1993Jaldering,2004Maund}, SN~2011dh \citep{Maund2011, 2011dh_image, arcavi11, Bersten2012, Folatelli2014}, and SN~2013df \citep{2013df}.
%One of the principal and fundamental properties of SN IIb is the disappearing H- lines in two month after the explosion, which monotonically the He-lines be performed the dominates lines in the spectra. The prevalent substantiate theory is fast expansion of the hydrogen envelope allowing the appearing of the deeper layer as a He-lines (these massive stars formed the he-core).

In some rare cases, massive stars may lose their hydrogen-rich envelope of a few solar masses and explode as SNe IIb with a prominent early flux excess that precedes the main radioactive peak \citep[e.g.,][]{arcavi11,Sana+12,GalYam2017,Fang+22}.
The prominent early peak in the optical light curve can last for a few days after explosion. This peak is believed to come from the cooling of shock-heated ejecta after shock breakout and is thus called a shock cooling light curve (SCL; \citealp{GalYam2017}). These double-peak phenomena have been discovered and analyzed in a few cases such as SN~1993J \citep{1993Jspectra, 2004Maund}, SN~2011dh \citep{arcavi11, 2011dhergon}, SN~2016gkg \citep{arcavi2017}, and SN~2017jgh \cite{ArmstrongTucker2021} etc.
For other multipeak SNe see \cite{Foley+07,Arcavi+17N,Gomez+21,zenati+22,chenz+22}

Continuous, high-cadence monitoring of the early light curve is key to capturing and analyzing the complete SCL. High-cadence imaging from space telescopes such as the \textit{Kepler Space Telescope} (\kepler; \citealp{haas2010kepler, Howell2014}) and the \textit{Transiting Exoplanet Survey Satellite} (\tess; \citealp{ricker2014}) are ideal to monitor such short-timescale transient phenomena. Observations from these telescopes have enabled some ground-breaking discoveries on the progenitors of various SNe \citep[e.g.,][]{dimitriadis2018k2,fausnaugh2019early,vallely2019, 2018agk, 2019esa, 2018lab}. In particular, 
\cite{ArmstrongTucker2021} analyzed the SCL of the Type IIb SN~2017jgh that was fully covered by \ktwo. With the high cadence coverage of the complete SCL, the progenitor properties were estimated with high precision. \cite{ArmstrongTucker2021} further demonstrate that without the high cadence \ktwo\ light curve during the rise, the fitting results would exhibit a systematic offset.

\begin{figure*}[t!]
    \centering
    \includegraphics[width=0.98\textwidth]{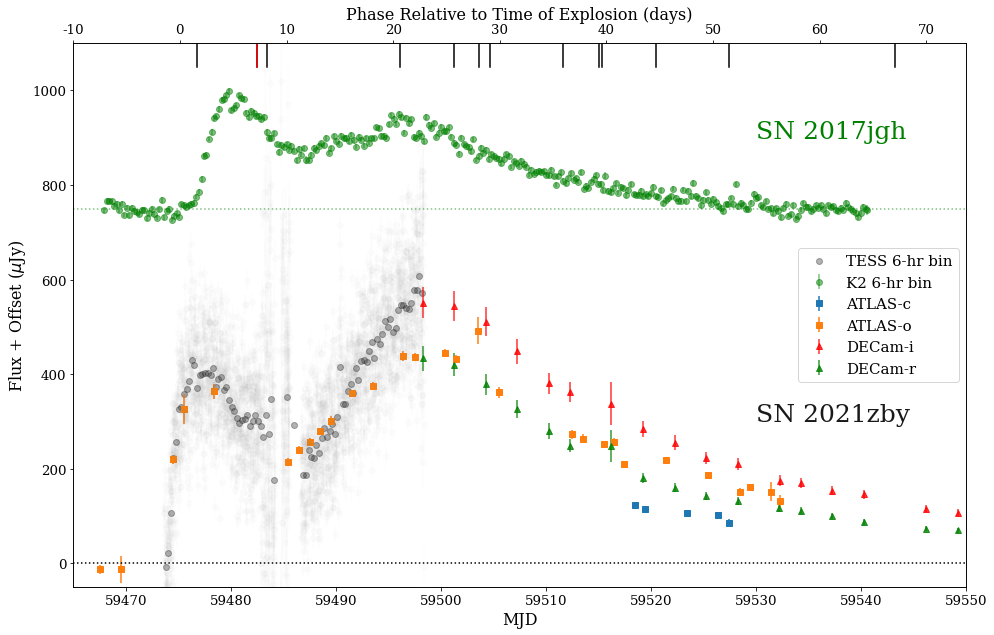}
    \caption{Optical light curves of SN~2021zby from \tess\ and ground-based surveys in comparison with the \textit{Kepler} light curve of SN~2017jgh \citep{ArmstrongTucker2021}. Black ticks at the top of the plot mark the times where ground-based optical spectra were obtained for SN~2021zby; red tick marks represent the NIR spectrum taken by IRTF. Notice that the \tess\ Sector 43 started $\sim$ 12 hour before the time of explosion and thus there's only one non-detection in the binned light curve in the pre-explosion phase.
%    \textcolor{magenta}{SJS: should plot the last non-detections The one from TESS at -12hr and the last one or two from ATLAS on 59469.5 and 59467.5}
}
    \label{fig:lightcurve}
\end{figure*}

Here we present the evolution of SN~2021zby during the first $\sim 2$ months after explosion with a spectro-photometric time series in the optical and near-infrared (NIR). SN~2021zby was discovered by the Asteroid Terrestrial-impact Last Alert System (ATLAS; \citealp{tonry2018atlas,atlasTransientServer}) on 2021 Sep 17 10:52:19.200 UTC, Modified Julian Day (MJD) 59474.45, in the $o-$band with $m_o = 18.16\pm0.14$ \citep{atlas_disc}, and was spectroscopically classified as an SN~IIb \citep{classification1,classification2}. 

SN~2021zby is located in a spiral arm of NGC~1166 at coordinates $\alpha=\rm 03^h00^m35.63^s$, $\delta= +11{\degr}50{\arcmin}29.74{\arcsec}$ (J2000.0). Prior to SN~2021zby, two other transients, the unclassified PS1-14abm \citep{PS1DR} and SN II 2018htf\citep{2018htfdic, 2018htfclas}, had been discovered in NGC~1166 in the past 10 years. Both of them are at least 3\arcsec\  away and are not associated with SN~2021zby.
%\redpen{PS1-14abm? are they associated? Seems not.} 
We adopt a redshift of $z = 0.025965\pm0.00002$ from \ion{H}{1} 21 cm measurements \citep{2005ApJS..160..149S} and a distance of 106.1 Mpc, corresponding to a distance modulus of $\sim 35.12$ mag. The Milky Way extinction is relatively high towards this direction, with $E(B-V)_{MW} = 0.21$ \citep{2011ApJ...737..103S}. The early \tess\ light curve of SN~2021zby starts at MJD 59473.8 and ends at MJD 59498.4, shortly before the time of the main radioactive peak. Combined with the DECam$-i$ and ATLAS$-o$ measurements with similar effective wavelengths around similar phase, we can infer the radioactive peak to be at $t_{max} = 59499\pm1$ MJD in \tess\ band. 

\tess\ coverage of SN 2021zby started shortly before explosion with $\sim 12$ hours of non-detection with a magnitude limit $m_{TESS}\gtrsim19.38$ at the beginning of sector 43. As inferred from the SCL fitting discussed in Section~\ref{sec:fit}, the time of explosion $t_0$, is around MJD $\sim 59474.4\pm0.1$, $\sim 24$ days prior to $t_{max}$. There was no clear detection of a short-duration shock breakout flash in the \tess\ light curve around the time of explosion. Systematic noise from scattered light and relatively low luminosity in redder bands like \tess\ may limit the detectability of the shock breakout flash in this case. Throughout this paper, phases are presented relative to the inferred time of explosion $t_{0}$, except for the model fitting section where $t_0$ is a free parameter to be constrained.

Throughout this paper, observed times are reported in MJDs while phases, unless noted otherwise, are reported in the rest-frame. We adopt the AB magnitude system, unless noted otherwise, and a flat $\Lambda$CDM cosmological model with $H_0 = 73$ km s$^{-1}$ Mpc$^{-1}$ \citep[][]{riess2016, riess2018}. All the data presented in this paper will be made public via WISeREP \footnote{https://www.wiserep.org/object/19385}\citep{wiserep}.

%\redpen{Previous studies on other CCSNe progenitor, direct images(93J, 11dh, maybe 87A and 05gl), SCL (11dh maybe, 17jgh)}

%11dh: directimage likely YSG, g'-lc matches $4 M_\odot$ He core, 270 $R_\odot$, $M_{e} = 0.1M_\odot$ and He mass fraction of envelope $=0.8$

\section{Observations and Data Reduction} \label{sec:data}

%from dust map in python I got E(B-V) = 0.244

%We've got photometric data from ATLAS through all phases, TESS in the rising phases and DECam\&PS1 in the post-peak phase. 

ATLAS has observed SN~2021zby since the pre-explosion stage in the $o-$band and during the post-peak stage in both $o-$ and $c$-band. We also obtained ground-based photometric follow-up in the post-peak stage with DECam on the CTIO 4-m Blanco telescope \citep{decam2008, 2015AJ....150..150F} in the $r$ and $i$ bands. Multi-band light curve is plotted in Figure~\ref{fig:lightcurve}, in comparison with the \kepler\ light curve of sn 2017jgh. In addition, we obtained spectra from multiple ground-based observatories. Details of these spectra are listed in Appendix.

To measure significant SN flux detection at the location of SN 2021zby, we apply several cuts on the total number of individual as well as averaged data in order to identify and remove bad measurements. Our first cut uses the chi-square and uncertainty values of the PSF fitting to clean out bad data. We then obtain forced photometry of 8 control light curves located in a circular pattern around the location of the SN with a radius of 17\arcsec. The flux of these control light curves is expected to be consistent with zero within the uncertainties, and any deviation from that would indicate that there are either unaccounted systematics or underestimated uncertainties. We search for such deviations by calculating the $3\sigma$-clipped average of the set of control light curve measurements for a given epoch (for a more detailed discussion see Rest et al, in prep.). This mean of the photometric measurements is expected to be consistent with zero and, if not, we flag and remove those epochs from the SN light curve. We then bin the SN 2021zby light curve by calculating a $3\sigma$-clipped average for each night, excluding the flagged measurements from the previous step. This method has been applied in a few other studies and proven its reliability in successfully removing bad measurements from the SN light curve \citep[e.g.][]{2020tlf}.

Following standard calibrations (bias correction, flat-fielding, and WCS) using the NSF NOIRLab DECam Community Pipeline \citep{valdes_decam_2014}, we reduced the DECam data using the \texttt{Photpipe} pipeline as described in \cite{rest_testing_2005, rest_cosmological_2013}. The images were warped into a tangent plane of the sky using the \texttt{swarp} routine \citep{bertin_terapix_2002}, after which photometry of the stellar sources was obtained using the standard PSF-fitting software DoPHOT \citep{schechter_dophot_1993}. We then used the PS1 catalog \citep{PS1catalogue} converted into the DECam natural system as described in \cite{Scolnic_supercal} to obtain the photometric zeropoints. The images were then kernel- and flux-matched to template images, subtracted, and masked using the \texttt{hotpants} code \citep{HOTPANTS}, which is based on the Alard-Lupton algorithm \citep{alardluptonref}. The 3-sigma clipped average position of SN~2021zby was calculated from all of the significant detections in the difference images to within 0.75\arcsec of the reported position. The final light curves were then obtained by performing forced photometry on this position for all images. The $r$- and $i$-band template images were taken after SN~2021zby was discovered, and therefore contain some residual SN flux. Thus, for each of these filters, we adjusted the light curves by adding a common flux offset so that the magnitudes match the ATLAS light curves, using an approximate conversion from \cite{tonry2018atlas}:%PS1?
%Firstly the ATLAS light curves in both bands are interpolated by linear functions in between MJD 59515 and 59535. Then we use the approximate conversion relation between ATLAS bands and $ri$ \cite{tonry2018atlas}
\begin{equation}
    r \sim 0.35\ c + 0.65\ o\ , \ i \sim -0.39\ c + 1.39\ o
\end{equation}

\begin{figure}[t!]
    \centering
    \hspace*{-0.1in}
    \includegraphics[width=0.41\textwidth]{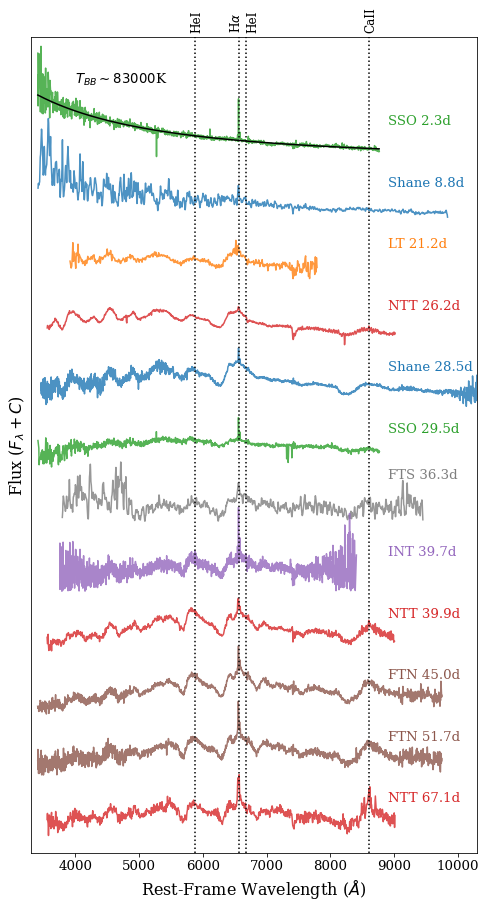}
    \hspace*{-0.1in}
    \includegraphics[width=0.41\textwidth]{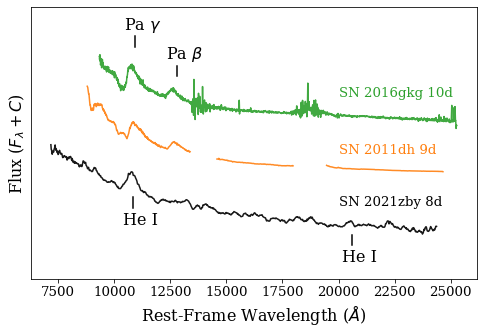}
    \caption{\textbf{Top:} Optical spectral series of SN~2021zby with phases and telescopes labeled above. Spectra taken with different telescopes are plotted with different colors. \textbf{Bottom:} NIR spectrum of SN~2021zby taken by IRTF around shock cooling peak, in comparison with SN~2011dh \citep{2011dhergon} and SN~2016gkg \citep{2016gkgTartaglia} around similar phase. The phases relative to the peak and relevant lines are labeled on the spectra.
    All the spectra have been normalized and shifted vertically for display purposes. }%\redpen{need to include 2016gkg}}
    \label{fig:spec}
\end{figure}

\tess\ observed SN~2021zby in Full Frame Images (FFIs) for sectors 42, 43, and 44 (S42, S43 and S44 thereafter), at a 10-minute cadence\footnote{The original calibrated \tess\ FFIs can be found in the the Mikulski Archive for Space Telescopes (MAST): \dataset[10.17909/0cp4-2j79]{http://dx.doi.org/10.17909/0cp4-2j79}}. These three sectors covered pre-explosion (S42), double-peak rise (S43), and decline (S44), which gives excellent coverage of the event outside of the 1~day mid-sector and inter-sector gaps. We reduce all sectors of TESS data using the \texttt{TESSreduce} python package, which aligns images, subtracts the variable background, and provides a flux calibration from field stars \citep{tessreduce}. One alteration was made to the default \texttt{TESSreduce} reduction where we included the nearby bad column 1167 in sector 43 data into the automatically determined strap mask. This inclusion rescaled the column producing a clean background subtraction for the nearby pixels that were used in the 3x3 aperture for SN~2021zby. Since \texttt{TESSreduce} produces difference-imaged light curves, we must add an offset to the light curves for S43 and S44 as the flux from SN~2021zby was included in the reference images. We estimated the offsets by calculating synthetic photometry in the TESS bandpass from spectra and photometry in other bands that are covered by TESS observations. Since the spectra do not cover the full TESS wavelength range $5802.57-11171.45$\AA, we extrapolate the spectra by assuming black body emission with temperature $\sim 83,000$ K estimated from the optical spectrum \footnote{Notice that this estimation can be biased because the optical and NIR spectra only covers the Rayleigh-Jeans tail and thus may not reflect the temperature of photosphere.} when calculating synthetic photometry. During S43 scattered light from the Earth and Moon in the detector reached saturation, making the photometry unreliable from MJD 59499 to 59524. 
%\textcolor{magenta}{SJS: that is very high. First, I don't think the shape in Fig 2 is 83,000K, and second you are correct that this is the tail and not reliable (are you sure it's not 8300K?).  High temperatures will be difficult to distinguish You should quote the lowest temperature that is compatible with this SED.}

\begin{figure*}[t!]
    \centering
    \hspace*{-0.1in}
    \includegraphics[width=\textwidth]{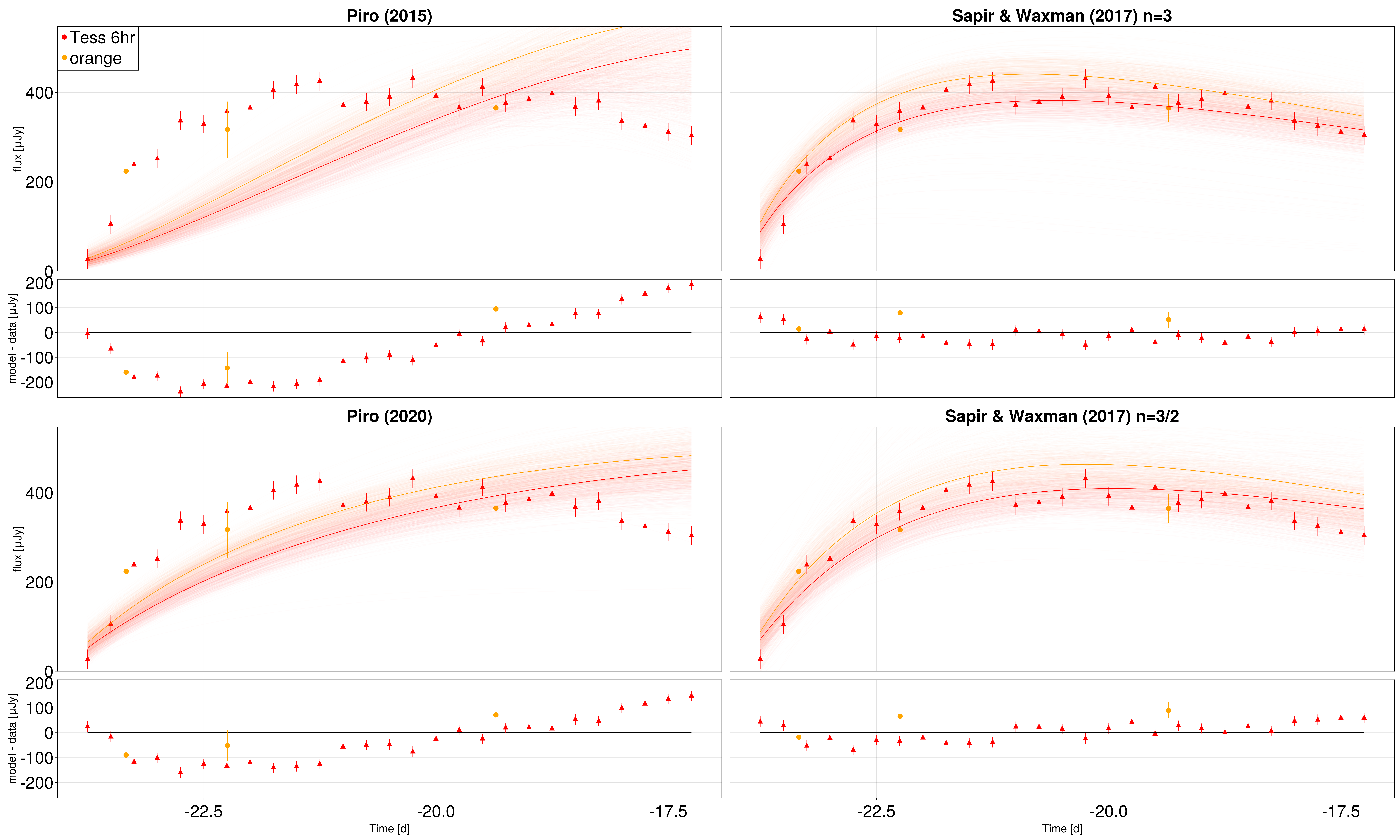}
    \caption{Model fits for the \tess\ and ATLAS-$o$ light curve of SN~2021zby during the shock cooling phase. Partially transparent lines are the 500 samples randomly drawn from the MCMC chain in order to visualize the posterior distributions. Time is relative to inferred \tess\ maximum. 
%    \textcolor{magenta}{SJS: why use the unbinned data points in ATLAS here, when the binned points are shown in Fig 1 and the statistical method for binning is well described ? We are arguing in Section 2 that this is the best method to treat the ATLAS data, but then don't use it. Why are the TESS points here different to Figure 1 ? It looks like they are 1hr binned points but Fig 1 has "6hr bins", so why is Fig 1 lightcurve noisier than Fig 3 ?}
    }
    \label{fig:fit}
\end{figure*}

%\redpen{spectra and spectra reduction. }
We obtained optical spectra with instruments including: the Wide-Field Spectrograph (WiFeS) \citep{WIFESinstrument} on the 2.3m telescope at Siding Spring Observatory (SSO), the IDS long-slit spectrograph on the 2.5m Isaac Newton Telescope (INT), the Kast spectrograph on the 3-m Shane Telescope at Lick Observatory, the ESO Faint Object Spectrograph and Camera \citep[EFOSC2,][]{EFOSC2} on the ESO New Technology Telescope (NTT, as part of the ePESSTO+ survey, \citealt{Smartt15AA}), the Spectrograph for the Rapid Acquisition of Transients (SPRAT; \citealt{Piascik2014}) on the Liverpool Telescope, and the Las Cumbres Observatory FLOYDS spectrographs mounted on the 2-meter Faulkes Telescope North (FTN) and South (FTS) at Haleakala Observatory and SSO, respectively, through the Global Supernova Project.

The IDS spectrum was reduced and flux-calibrated appropriately using the standard the Image Reduction and Analysis Facility (IRAF; \citealp{tody86}) \texttt{specred} routines.
The EFOSC2 spectra were reduced in a similar manner, with the aid of the PESSTO pipeline\footnote{\url{https://github.com/svalenti/pessto}}. 
The KAST spectra were reduced using a custom data reduction pipeline based on IRAF.\footnote{The pipeline is publicly accessible at \url{https://github.com/msiebert1/UCSC_spectral_pipeline}.}
One-dimensional FLOYDS spectra from FTN and FTS were extracted, and flux and wavelength calibrated using the \texttt{floyds\_pipeline}\footnote{\url{https://github.com/LCOGT/floyds_pipeline}} \citep{Valenti2013}.
The WiFeS data were processed with the PyWiFeS pipeline\footnote{\url{https://www.mso.anu.edu.au/pywifes/doku.php}} \citep{PYWIFES}.

We obtained a near-IR spectrum of SN~2021zby on 2021 Sep 25 using the SpeX spectrograph \citep{rayner2003} on the NASA InfraRed Telescope Facility (IRTF). 
We used the low-resolution prism mode with the 0.8$''$ slit, providing a resolving power of $R\sim 75$ with a simultaneous coverage between 0.7--2.5 $\mu$m. 
We observed an A0V star HIP16095 immediately after the SN for telluric correction and flux calibration. 
We reduced the data using \texttt{spextool} \citep{cushing2004}, which performed flat field correction, wavelength calibration, and spectral extraction.
Telluric correction was performed using \texttt{xtellcor} \citep{vacca2003}. The optical and NIR spectral series of SN~2021zby are shown in Figure~\ref{fig:spec}.

%\begin{figure}[t!]
%    \centering
%    \caption{optical spectra}
%    \label{fig:spec_optical}
%\end{figure}

\section{analysis} \label{sec:style}

%extinction: host extinction\\

%\redpen{pseudo-bolo luminosity estimate and compare with other SNe with SCL}

\begin{figure*}[t!]
    \centering
    \hspace*{-0.1in}
    \includegraphics[width=1\textwidth]{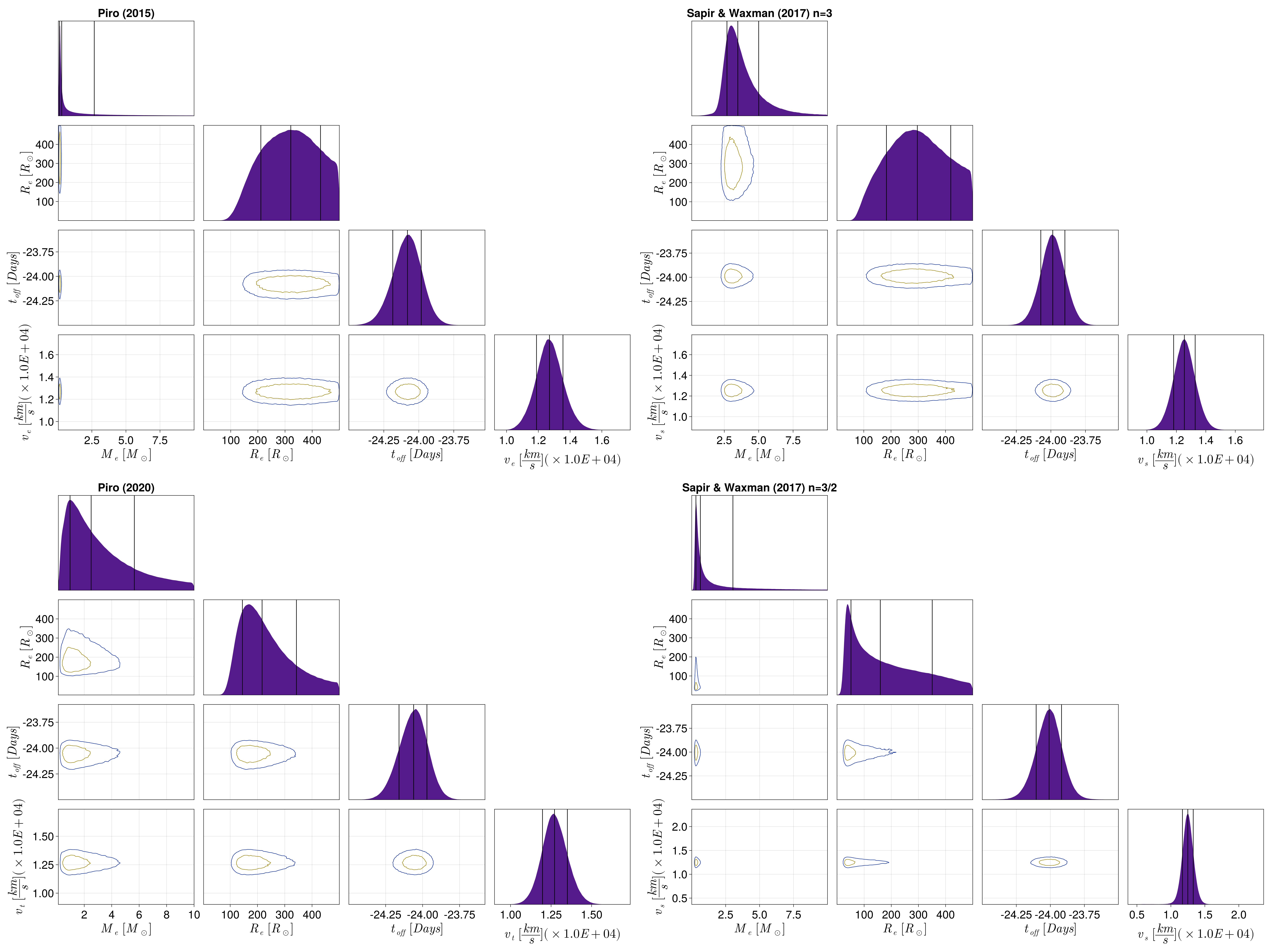}
    \caption{Corner plots of our light curve fit to SN 2021zby. The brown line outlines the $1-\sigma$ joint posterior. The three lines in the individual posterior plots mark the 16th, 50th and 84th percentiles of parameter posteriors, taken as lower-bounds, best fitting values, and upper-bounds. Note that the non-Gaussian posterior may cause biases in determining best fitting values.
%    \textcolor{magenta}{Hard to read these axes labels, small and also very faint}
    }
    \label{fig:corner}
\end{figure*}

\begin{deluxetable*}{cccccc}
\centering 
\tablecolumns{6}
%\tablewidth{1.8*\textwidth} 
\tablecaption{Priors and results of shock cooling models fit to the shock cooling light curve of SN~2021zby. The upper- and lower-bounds are determined by taking 16th and 84th percentiles of the parameter posteriors. The $M_{e}$ prior was chosen based on physical expectations and is constrained to have a maximum of 10~$M_{\odot}$. Similarly, the prior for $R_{e}$ was chosen based on physical expectations. The $v$ prior was chosen from constraints calculated via a Gaussian fit to He I $10830\AA$ around the shock cooling peak. Finally the constraint on $t$ was chosen from the position of the first \tess\ observation. The Log-likelihood of each model fit is also listed.}
\tablehead{\colhead{Parameter} & 
\colhead{Envelope Mass($M_{e}$)} & 
\colhead{Envelope Radius ($R_{e}$)}  & 
\colhead{Velocity($v$)}& 
\colhead{Start time ($t$)}& 
\colhead{Log-likelihood}}
\startdata
Prior Function& Levy & Uniform & Normal& Normal & - \\
PriorInputs &  [Location, Scale] & [Min, Max] & [Mean, Std] & [Mean, Std] & - \\
Input Values & [0, 10]$M_{\odot}$ & [0, 500]~$R_{\odot}$ & [12500, 800]km/s& [-24, 0.5]~days & -\\
SW17 $n=3/2$ & $0.638^{+2.396}_{-0.328}M_\odot$ & $160^{+191}_{-108} R_\odot$ & $12491^{+798}_{-790}$km/s & $-24.009^{+0.091}_{-0.094}$days &-14.88\\
SW17 $n=3$ & $3.483^{+1.517}_{-0.789}M_\odot$ & $297^{+123}_{-113} R_\odot$ & $12542^{+734}_{-728}$km/s & $-23.986^{+0.088}_{-0.086}$ days & -16.26 \\
P15 & $0.287^{+2.390}_{-0.158}M_\odot$ &  $321^{+109}_{-110} R_\odot$ & $ 12703^{+850}_{-818}$km/s& $-24.081^{+0.099}_{-0.105}$ days & -34.29 \\
P20 & $2.503^{+3.146}_{-1.537}M_\odot$ & $217^{+126}_{-72} R_\odot$ & $12712^{+793}_{-741}$km/s &   $-23.986^{+0.090}_{-0.099}$ days &   -23.42\\
\enddata
\label{tab:prior}
\end{deluxetable*}
\subsection{Fitting light curve with models}\label{sec:fit}
%\redpen{Patrick: can you roughly describe the new model fitting here?}
There are a number of semi-analytical shock cooling light curve models available, including~\citet{Piro2015} (hereafter P15),~\citet{PiroHaynie2020} (hereafter P20), and~\citet{SapirWaxman2017} (hereafter SW17). P15 is the simplest of these, making no assumption about the density profile of the progenitor and assuming a simple expanding photosphere. P20 is a revision of the P15 model which improves upon P15 by employing a two-component velocity model. P20 models the progenitor with outer material which has a steep velocity gradient, and inner material with a shallow velocity gradient. SW17 assumes the progenitor has a polytropic density profile. This is characterised by the polytropic index $n$ which is equal to $3/2$ for progenitors with a convective envelope, such as red supergiants (RSG) or yellow supergiant (YSG), and equal to $3$ for progenitors with a radiative envelope, such as blue supergiants (BSG). Each model is parameterised by an envelope mass ($M_{e}$), envelope radius ($R_{e}$), velocity ($v$), and start time ($t$) which is relative to the peak of the radioactive portion of the light curve. 

Each of these models assume that the progenitor radiates as a black body and uses an analytical description of the luminosity, radius, and temperature of the progenitor over time in order to derive the flux of the shock cooling light curve. Full details of these analytical models can be found in~\cite{ArmstrongTucker2021}, who we closely follow.

Following \cite{ArmstrongTucker2021}, we fit the shock cooling models using an affine-invariant MCMC method (ShockCooling.jl: \url{https://github.com/OmegaLambda1998/ShockCooling}). This algorithm produces an approximate posterior for a model, given data, priors, and a likelihood function. Our data consists of the ATLAS-$o$ and the 6 hour binned \tess\ light curve, up to$\sim 7$ days after explosion, the time when the shock cooling light curve transitions into the radioactive light curve. Our likelihood function is chosen to be the reduced $\chi^2$ between the models and our data. Both bands are fitted simultaneously, and the combined reduced $\chi^2$ is then minimised. The reduced $\chi^2$, defined as $\chi^2$ divided by the degrees-of-freedom of each band, allows us to weight each reduced $\chi^2$ by the number of data points, accounting for the larger sample of \tess\ data. 
Due to the degeneracy inherent in the models, the choice of priors has a large effect on the final fit. As such, we used an iterative approach to defining the priors, starting with large uniform priors with $0<R_e/R_\odot<500$ and $0<M_e/M_\odot<10$ and then using the posterior to update our choice of prior. The prior of ejecta velocity is constrained to be $12500\pm800$ km/s, determined by the FWHM of the $1.1~\mu$m feature around the shock cooling peak as discussed in the following section. Our final priors for each parameter are listed in Table 1. The best-fit values of SW17 models satisfy the validity range of temperature $T>0.7$eV within the fitting range.

The fitting results for four models discussed are also included in Table 1 and plotted in Figure~\ref{fig:fit}. Figure~\ref{fig:corner} shows the corner plots of posterior distributions of each parameter. The lower-bound, best-fitted value, and upper-bound are determined by the 16th, 50th, and 84th percentiles in the posterior distributions.
%Among these models, the log-likelihood for SW17 RSG and BSG models are -14.88 and -16.26 respectively, while the Piro+15 and Piro+20 models have significantly larger residuals and their log-likelihood are -34.29 and -23.42 respectively. 
Among these models, SW17 $n=3/2$ and $n=3$ models have significantly smaller residuals and larger log-likelihood than P15 and P20 models.
Thus, from the light curve fit alone, the SW17 $n=3/2$ models are preferable in general, while the SW17 $n=3$ model performs marginally worse than the $n=3/2$ model. We note that the posteriors of some parameters are highly non-Gaussian, e.g. $R_e$ in SW17 $n=3/2$ model, and thus may not accurately represent the best-fit values. For future analyses, we will explore refinements to our SCL-fitting procedure such as using a redefined maximum likelihood estimator, which could possibly improve the P15 and P20 fits.

%\redpen{old values, need change} Among the four models we found that SW17 BSG model gives the best fit to the shock cooling light curve in multiple bands. The best-fit model and data are shown in Figure~\ref{fig:fit} along with residuals. From the fitting we estimated $M_e = 18.87^{+6.07}_{4.55} M_\odot$ and $R_e = 48.40^{+9.18}_{-8.64}R_\odot$. This mass is  The ejecta velocity is $12347.14^{+738.50}_{-728.63}$ km/s, closely matches the value we measured from the spectrum around peak. 

%Acceptance ratio: 0.57
%M: 18.87 + 6.07 / -4.55 Solar masses
%R: 48.40 + 9.18 / -8.64 Solar radii
%Explosion Time: -23.97 + 0.089 / - 0.089 days relative to radioactive peak
%Velocity: 12347.14 + 738.50 / -728.63 km / s

%Rabinak \& Waxman (2011) (see their eq. 12) for radius estimation

%dark phase?\\

%Companion? maybe worth a rough estimation on energy scale to see if this can produce prominent bump (around first peak).\\

\subsection{Spectroscopic features}\label{section:spec}

\begin{figure*}[t!]
    \centering
    \hspace*{-0.1in}
    \includegraphics[width=1\textwidth]{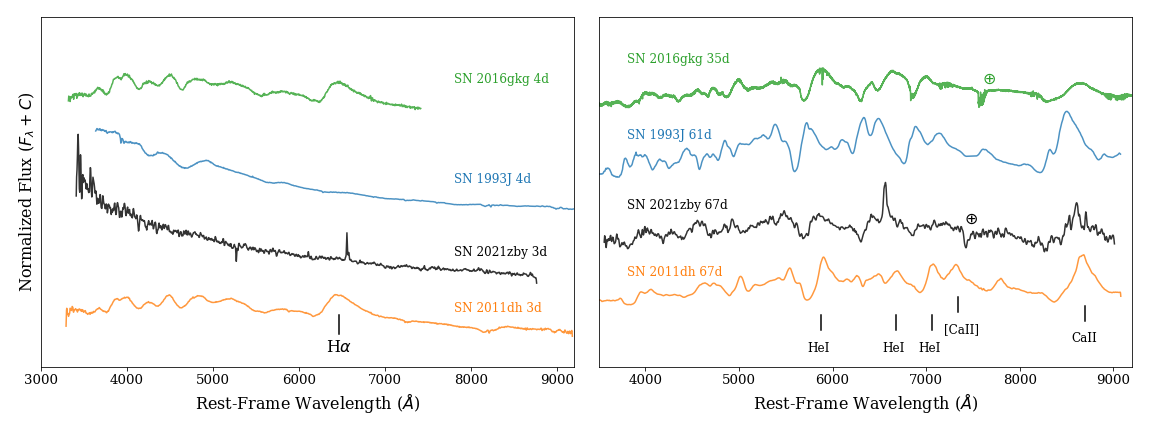}
    \caption{The spectra of SN~2021zby on 3 days (left) and 67 days (right) after explosion, in comparison with SN~2016gkg \citep{2016gkgTartaglia}, SN~2011dh \citep{2011dhergon} and SN~1993J \citep{1993Jspectra}. All the spectra have been normalized and shifted for clarity.
    }
    \label{fig:spec_comp}
\end{figure*}

As shown in Figure~\ref{fig:spec}, the optical spectra around the shock cooling phase are dominated by black body continuum with $T_{BB}\gtrsim10^4$ K with few line features except the narrow H$\alpha$ emission from the host. In Figure~\ref{fig:spec_comp}, we further include the optical spectra from other SNe~IIb with SCL around shock cooling phase and late phase. Around the shock cooling peak, there is no detection of broad H$\alpha$ emission in the spectra of SN~2021zby, which is a prominent feature in SN~2011dh and SN~2016gkg at a similar phase \citep{arcavi11, 2016gkgTartaglia}. \cite{1994Jnarrow} reported the discovery of narrow lines with $v < 1000$km/s in SN~1993J around the shock cooling phase, including H$\alpha$, He II, [Fe X] and [Fe XIV], and claimed that those narrow emission lines are signals of circumsteller medium (CSM) interaction. Such features are not seen in the early spectra of SN~2021zby, possibly due to the relatively low S/N and spectral resolution. As shown in Figure~\ref{fig:spec} and~\ref{fig:spec_comp}, the optical spectra after the shock cooling phase during the radioactive peak start to resemble the spectral evolution of a typical SN~IIb. 

The weak H$\alpha$ feature of SN 2021zby is persistent and still observable even at 43 days after the radioactive peak. This is different from SN~1993J for which the H$\alpha$ feature significantly weakened at a similar phase \citep{1993Jspectra}. This may indicate a sizeable mass of Hydrogen in the progenitor of SN~2021zby. 

%\redpen{Luc's comment: At those early times when the spectra are very blue, H may be partially ionized, so one may see a weak Halpha and HeI10830. The weakness of hte lines may also be related to the density structure, with RHO falling very steeply. As times passes, ionization drops and HI lines strengthen (Halpha is quite strong and persistent). Then, in the NIR, the HeI10830 line should still be there but the HI lines should be present too. What is surprising in your observations is that Halpha is there at all times, even at 43d. Clearly, there is a sizeable H mass in this object. You could compare to the evolution seen in 93J. Also, have a look at my 2018 paper (as well as older ones in 2011 and 2015 where I show some results for SNe IIb) }

%1993J:  There are some broad undulations that may be incipient P-Cygni features of $H-\alpha$ and He I $\lambda5876$, but the presence of reduction artifacts makes interpretation problematic. The day 3 spectrum also contained narrow (unresolved) emission features of He II $\lambda4686$, [Fe X] $\lambda6374$, and $H-\alpha$.

The NIR spectrum taken by IRTF around the shock cooling peak is plotted in the bottom panel of Figure~\ref{fig:spec} in comparison with NIR spectra of SN~2016gkg and SN2011dh around similar phase. The only significant feature is the broad emission line at around $10800$\AA, which may come from Pa$\gamma$, He I $1.083\mu$m, C I $1.0693\mu$m, or Mg II $1.0927\mu$m as discussed in \cite{CSP-SECCSN2022}. Unlike SN 2011dh and SN 2016gkg, there is no detection of Pa $\beta1.2818\mu$m feature. On the other hand, a weak emission feature is present around He I $2.0581\mu$m, while there are no detections of Mg II $2.1369\mu$m in the NIR or any other H I or C I features in the optical spectrum at a similar phase. We therefore conclude that this $1.1\mu$m feature most likely comes from He I $1.083\mu$m. We measure the FWHM of this $1.1\mu$m feature with a simple Gaussian fit and use it as the prior of the ejecta velocity in the light curve fitting (see Table~1).

\section{Discussion and Conclusion}

%The high cadence light curves from \kepler\ and \tess\ are a unique window to reveal the complete shock cooling light curves of SNe IIb. 
The progenitor of SN~2021zby is likely to have a moderately extended envelope to produce a clear shock cooling peak, but not as extended as those of SNe~IIP progenitors as the peak is not blended with the radioactive peak. With the high-cadence \tess\ and ATLAS$-o$ light curves covering the full shock cooling phase of SN~2021zby, we are able to constrain the progenitor's properties with relatively high precision compared to ground-based observations alone. We fit the multi-band light curves following the fitting scheme described in \cite{ArmstrongTucker2021},
%with ejecta velocity constrained by the broad \ion{He}{1} ($\lambda$ 10830) spectral feature near the time of the shock-cooling peak. 
%For SW17 RSG model, the fitting results indicate the envelope has a mass of $M_e = 0.638^{+2.396}_{-0.328} M_\odot$ and a radius of $R_e = 159.535^{+191.302}_{-107.956} R_\odot$, and ejecta velocity is $v_s = 12491.497^{+797.953}_{-789.838} $km/s. The time of explosion is $t_{off} = -24.009+0.091/-0.094 d$ relative to the atlas-$o$ band peak. For BSG model, the fitted parameters are $M_e =  3.483^{+1.517}_{-0.789} M_\odot$, $R_e = 296.653^{+122.569}_{-113.425} R_\odot$, $t_{off} = -23.986+0.088/-0.086 d$, $v_s = 12491.497^{+797.953}_{-789.838} $km/s, respectively.
and the best-fit models are:
\begin{itemize}
    \item The SW17 $n=3/2$ model (convective envelope, i.e. similar to those of RSG or YSG) indicates a progenitor with envelope mass of $\sim 0.3-3.0 M_\odot$ and envelope radius of $\sim 50-350 R_\odot$. 
    \item The SW17 $n=3$ model (radiative envelope, i.e. similar to that of BSG) gives the second-best fit with a marginally larger log-likelihood, and indicates an envelope mass of $\sim 2.7-5.0 M_\odot$ and envelope radius of $\sim 184-420 R_\odot$. However, such an envelope radius is significantly larger than the expected radii of BSGs, which are in the range of $40-80R_\odot$\citep{Underhill1979}. This implies that it is physically inconsistent with observations of the structure of BSGs.
\end{itemize}
%Overall, the SW17 RSG model with $n = 3/2$ gives the best fit to the multi-band light curve of SN~2021zby, indicating a progenitor with envelope mass of $\sim 0.3-3.0 M_\odot$ and envelope radius of $\sim 50-350 R_\odot$. We still cannot fully exclude the possibility of a BSG progenitor with an envelope mass of $\sim 2.7-5.0 M_\odot$ and envelope radius of $\sim 11-420 R_\odot$ as the SW17 BSG model performs only marginally worse. 

In the previous study on SN~2017jgh, a similar fitting schemes was applied to its \kepler\ light curve \citep{ArmstrongTucker2021}, and they also found that the  SW17 $n=3/2$ model is the best-fit model, while the SW17 $n=3$ model performs marginally worse. The best-fit result of SN~2017jgh indicates its progenitor to be most likely a YSG with envelope radius $R_e \sim 50-290 R_\odot$ and envelope mass $M_e \lesssim 1.7 M_\odot$%,  \redpen{likely due to the relatively longer duration of SCL in SN~2021zby}
. A few other SNe IIb has progenitors discovered in pre-SN images, enabling another approach to constrain progenitor properties. \cite{1993Jaldering} identify the progenitor of SN~1993J as a YSG of type K0 Ia with SED and luminosity, and estimated the radius to be $\sim 500R_\odot$. For SN~2016gkg, \cite{2016gkg3} constrained the progenitor to be a YSG as well, with radius $\sim70R_\odot$, which matches $40-150R_\odot$ estimate from light curve fit by \cite{arcavi2017}. For SN~2011dh, \cite{Bersten2012} compared its $g$-light curve with models and also found a YSG with $R \sim 200R_\odot$ to be the best match for its progenitor, which is confirmed by the pre- and post-explosion images \citep{2011dh_image}.  The progenitor radius of SN~2021zby estimated from the best-fit SW17 $n=3/2$ model lies in between these SNe IIb confirmed with YSG progenitors, indicating that SN~2021zby may have a YSG progenitor as well, though the possibility of RSG progenitor cannot be excluded by current light curve fitting alone. %Note that the parametrized models used in this paper can be over-simplified, e.g. \textbf{specify the where}. Comparing with state-of-the-art non-parametrized simulation such as \textbf{specify what models}, it’s \textbf{describe bias here}.

%most likely yellow SG, fitted well to n=3/2(RSG) with $R = 128^{+160}_{-76}R_\odot$, $M_e = 0.56^{+1.17}_{-0.28}$, $v_s = 8.8^{+1.5}_{-1.3}\times10^3$km/s, $t_off = -21.04^{+0.25}_{-0.27}$, while BSG fit gives a fit with slightly larger chi2, and $R = 311^{+120}_{-98}R_\odot$, $M_e = 3.09^{+1.09}_{-0.75}$, $v_s = 8170^{+830}_{-160}$km/s, $t_off = -20.96^{+0.23}_{-0.21}$. 

%17jgh, most likely yellow SG, fitted well to n=3/2(RSG) with $R = 128^{+160}_{-76}R_\odot$, $M_e = 0.56^{+1.17}_{-0.28}$, $v_s = 8.8^{+1.5}_{-1.3}\times10^3$km/s, $t_off = -21.04^{+0.25}_{-0.27}$, while BSG fit gives a fit with slightly larger chi2, and $R = 311^{+120}_{-98}R_\odot$, $M_e = 3.09^{+1.09}_{-0.75}$, $v_s = 8170^{+830}_{-160}$km/s, $t_off = -20.96^{+0.23}_{-0.21}$. For SN 2011dh \cite{Bersten2012} compared $g'$-lightcurve with models and found $R \sim 200R_\odot$ to be the best match. \cite{2004Maund} .

%Similar fit with all four models has been applied to SN 2017jgh, and SW17 RSG model with $n = 3/2$ is also the most preferrable \citep{ArmstrongTucker2021}.

%2016gkg: also a YSG with R~ 30-150 with pre-SN image

%\redpen{Discuss issues on priori choice.}

%\redpen{Discuss constraints on the progenitor, size, pre-explosion activity etc.}

The early spectra of SN~2021zby during SCL shows clear differences compared to the other SNe IIb with similar SCLs. Unlike SN 2011dh and SN 2016gkg, SN~2021zby lacks H features in the shock cooling phase. Such a phenomenon might be a consequence of high ionization around the shock cooling phase \citep{Dessart2018}. This argument is further supported by the blue continuum of the early spectra. On the other hand, the presence of broad He I$10830$ at early times and a relatively strong H$\alpha$ feature after 43 days post-peak indicates that the progenitor still had a sizable amount of H and He. 

The high cadences of \kepler\ and \tess\ are crucial to sufficiently constrain the progenitor properties with the model fitting of the SCL. However, even with such exquisite data, it is still challenging to fully break the degeneracy between the RSG and YSG progenitors, which could be distinguished either through observations in the UV band, where the signal is most prominent, or by constraining the wind speed with the flash ionized features with high-resolution spectra. With improving cadence and receiving earlier alerts from, for example, Rubin Observatory \citep{LSST} in the near future, we can expect to have better time and wavelength coverage in bluer bands. In the longer term, the next generation of MIDEX space-based UV telescopes, e.g. UVEX\citep{uvex} and STAR-X \citep{STARX}, will allow us to monitor SCLs in the UV. 
%STAR-X\citep{STARX}
%In summary, this further validates the importance of the high cadence and early coverage from \kepler\ and \tess\ in making a good fit to SCL. Even with this data it is still challenging to fully break the degeneracy between SW17 RSG and BSG models. The early spectra of SN~2021zby during SCL shows clear difference comparing to other SNe IIb. Thus, fine modelling and more spectroscopic follow-up on the this phase might be very useful in further characterize the progenitors of stripped-envelope SNe (SESNe) with SCL. 

%\section{Conclusions}

%\redpen{Compare with previous examples, 93J?}

%Plots
%\begin{itemize}
%    \item full lightcurves 
%    \item maybe a plot comparng 93J and 16gkg and 11dh lightcurves?
%    \item spectra series and comparison with others
%    \item zoom into H-alpha fitting maybe
%    \item model fitting 
%    \item corner plots
%\end{itemize}

%page limit for a letter: For MNRAS, 5 page limit including reference, 200 words for abstract. For ApJ: 3500 words for manuscript, 250 for abstract, 50 references, 5 Figures \& Tables

\section*{Acknowledgement}

The author acknowledges Luc~Dessart, Eli~Waxman, Nir~Sapir, Ryosuke~Hirai and Sasha~Kozyreva for the valuable discussion.

This paper includes data collected by the TESS mission. Funding for the TESS mission is provided by the NASA's Science Mission Directorate. The \tess\ data presented in this paper were obtained from the Mikulski Archive for Space Telescopes (MAST) at the Space Telescope Science Institute (STScI). The specific observations analyzed can be accessed via \dataset[https://doi.org/10.17909/0cp4-2j79]{https://doi.org/10.17909/0cp4-2j79}. STScI is operated by the Association of Universities for Research in Astronomy, Inc., under NASA contract NAS5–26555. Support to MAST for these data is provided by the NASA Office of Space Science via grant NAG5–7584 and by other grants and contracts.

This work makes use of data from Las Cumbres Observatory.  The LCO group is supported by NSF grants AST-1911225 and AST-1911151.

This work was funded by ANID, Millennium Science Initiative, ICN12\_009.

The Isaac Newton Telescope is operated on the island of La Palma by the Isaac Newton Group of Telescopes in the Spanish Observatorio del Roque de los Muchachos of the Instituto de Astrofísica de Canarias (I/2021B/14). 

Q.W. acknowledges financial support provided by the STScI Director's Discretionary Fund.

I.A. is a CIFAR Azrieli Global Scholar in the Gravity and the Extreme Universe Program and acknowledges support from that program, from the European Research Council (ERC) under the European Union’s Horizon 2020 research and innovation program (grant agreement number 852097), from the Israel Science Foundation (grant number 2752/19), from the United States - Israel Binational Science Foundation (BSF), and from the Israeli Council for Higher Education Alon Fellowship.

The UCSC team is supported in part by NASA grant 80NSSC20K0953, NSF grant AST--1815935, the Gordon \& Betty Moore Foundation, the Heising-Simons Foundation, and by a fellowship from the David and Lucile Packard Foundation to R.J.F.

L.G. and T.E.M.B. acknowledge financial support from the Spanish Ministerio de Ciencia e Innovaci\'on (MCIN), and the Agencia Estatal de Investigaci\'on (AEI) 10.13039/501100011033 under the PID2020-115253GA-I00 HOSTFLOWS project, from Centro Superior de Investigaciones Cient\'ificas (CSIC) under the PIE project 20215AT016 and the I-LINK 2021 LINKA20409, and the program Unidad de Excelencia Mar\'ia de Maeztu CEX2020-001058-M.
L.G. additionally acknowledges the European Social Fund (ESF) "Investing in your future" under the 2019 Ram\'on y Cajal program RYC2019-027683-I.

M.G. is supported by the EU Horizon 2020 research and innovation programme under grant agreement No 101004719.

N.I. was partially supported by Polish NCN DAINA grant No. 2017/27/L/ST9/03221.

J.D.L. and D.O.N. acknowledge support from a UK Research and Innovation Fellowship (MR/T020784/1).

M.N. is supported by the European Research Council (ERC) under the European Union’s Horizon 2020 research and innovation programme (grant agreement No.~948381) and by a Fellowship from the Alan Turing Institute.

SJS, SS, DRY and KWS acknowledge funding from STFC Grants ST/T000198/1 and ST/S006109/1. 
 
\section*{Software and third party data repository citations} \label{sec:cite}


%\vspace{5mm}
\facilities{\tess, Shane, NTT, FTN, FTS, IRTF, DECam, ATLAS, INT, SSO:2.3m, LT}


\software{astropy \citep{2013A&A...558A..33A,2018AJ....156..123A},
\texttt{TESSreduce} \citep{tessreduce}, ShockCooling.jl: \url{https://github.com/OmegaLambda1998/ShockCooling},
Matplotlib \citep{matplotlib}, 
SciPy \citep[][]{2020SciPy-NMeth}, 
NumPy \citep{2020NumPy-Array}, 
pysynphot \citep[][]{2013ascl.soft03023S},          }



\appendix

\startlongtable
\begin{deluxetable*}{ccccc}
\centering
\tablecolumns{5} \tablewidth{1pc} 
\tablecaption{Log of Spectroscopic Observations of SN 2021zby}
\tablehead{\colhead{MJD} & \colhead{Phase relative to $t_{0}$} & \colhead{Phase relative to $t_{peak}$} & \colhead{Telescope/Instrument}  & \colhead{Wavelength Range } \\
 & [days] & [days] & & [{\AA}]}
\startdata 
59476.745 &  2.29 & -21.69 &  SSO 2.3m/WiFeS & 3500-9000\\
%59477.533 &   &  -20.92     SCAT-UH88/SNIFS & ?-?\\
59482.491  & 7.89 &  -16.09   &  IRTF/SpeX & 6848-25378\\
59483.466 & 8.84  &  -15.14   &  Shane/KAST & 3506-10094\\
59496.114 & 21.16  &  -2.81    &  LT/SPRAT & 4020-7994\\
59501.265 & 26.18  &   2.21    &  NTT/EFOSC2 & 3652-9248\\
59503.594 & 28.45  &   4.48    &  Shane/KAST & 3256-10896\\
59504.653 & 29.48  &    5.51   &  SSO 2.3m/WiFeS & 3500-9000\\
59511.619  & 36.28 &   12.30   &  FTS/FLOYDS & 3500-10000\\
59515.092 & 39.66  &   15.68   &  INT/IDS & 3855-8627\\
59515.300 & 39.86  &   15.89   &  NTT/EFOSC2 & 3651-9245\\
59520.520 & 44.95  &   20.98   &  FTN/FLOYDS & 3500-10000\\
59527.470 & 51.73  &   27.75   &  FTN/FLOYDS & 3500-10000\\
59543.206 & 67.06  &   43.08   &  NTT/EFOSC2 & 3652-9248 
\enddata
\label{tab:spec}
\end{deluxetable*}

\bibliography{reference}{}
\bibliographystyle{aasjournal}



\end{document}